\documentclass{revtex4}
\usepackage{amssymb,amsmath,amsfonts}
\usepackage{graphicx,graphics}
\usepackage{color}

\usepackage{framed}

\newcommand{\rem}[1]{}
\DeclareMathAlphabet{\mathbi}{OML}{cmm}{b}{it} 
\newcommand{\non}{\nonumber}
\newtheorem{theorem}{Theorem}
\newtheorem{lemma}{Lemma}

\newcommand{\bx}{\mathbi{x}}

\newcommand{\bel}{\begin{equation}\label}
\newcommand{\ee}{\end{equation}}
\newcommand{\beq}{\begin{eqnarray}\label} 
\newcommand{\eeq}{\end{eqnarray}} 
\newcommand{\bc}{\begin{center}} 
\newcommand{\ec}{\end{center}} 
\newcommand{\ben}{\begin{enumerate}}
\newcommand{\een}{\end{enumerate}}
\newcommand{\bit}{\begin{itemize}}
\newcommand{\eit}{\end{itemize}}
\newcommand{\I}{\int_{\mathcal{V}}}

\newcommand{\bdf}{\mathbi{f}}

\newcommand{\bu}{\mbox{\boldmath$u$}}

\newcommand{\bom}{\mbox{\boldmath$\omega$}}

\newcommand\shalf{\ensuremath{{\scriptstyle\frac{1}{2}}}}

\newcommand{\Gr}{Gr}

\makeatletter
\@addtoreset{equation}{section}

\makeatother

\begin{document}
\title{\sf\textbf{\color{blue}Conditional regularity of solutions of the three dimensional\\ 
Navier-Stokes equations and implications for intermittency}}
\author{J. D. Gibbon}
\affiliation{Department of Mathematics, Imperial College London SW7 2AZ, UK}
\email{j.d.gibbon@ic.ac.uk}
\homepage{http://www2.imperial.ac.uk/~jdg}

\begin{abstract}
Two unusual time-integral conditional regularity results are presented for the three-dimensional 
Navier-Stokes equations.  The ideas are based on $L^{2m}$-norms of the vorticity, denoted 
by $\Omega_{m}(t)$, and particularly on $D_{m} = \Omega_{m}^{\alpha_{m}}$, where 
$\alpha_{m} = 2m/(4m-3)$ for $m\geq 1$. The first result, more appropriate for the unforced 
case, can be stated simply\,: if there exists an $1\leq m < \infty$ for which the integral condition 
is satisfied ($Z_{m}=D_{m+1}/D_{m}$)
$$
\int_{0}^{t}\ln \left(\frac{1 + Z_{m}}{c_{4,m}}\right)\,d\tau \geq  0\,,
$$
then no singularity can occur on $[0,\,t]$. The constant $c_{4,m}\!\searrow \! 2$ for large $m$.
Secondly, for the forced case, by imposing a critical \textit{lower} bound on $\int_{0}^{t}D_{m}\,d\tau$, 
no singularity can occur in $D_{m}(t)$ for \textit{large} initial data.  Movement across this critical 
lower bound shows how solutions can behave intermittently, in analogy with a relaxation oscillator. 
Potential singularities that drive $\int_{0}^{t}D_{m}\,d\tau$ over this critical value can be 
ruled out whereas other types cannot.
\end{abstract}
\maketitle

\section{Introduction}\label{intro}

The twin related themes of this paper are firstly the regularity problem for solutions of the 
three-dimensional incompressible Navier-Stokes equations and secondly the intermittent 
behaviour of these solutions. Traditionally the first problem, which still remains 
tantalizingly open, has lain in the domain of the analyst, whereas the phenomenon of 
intermittency has tended to be more of interest to the physics and engineering fluid 
dynamics communities. It will be demonstrated in this paper that these two issues are 
intimately related \& require simultaneous study.  

\subsection{History}\label{hist}

Formally, a weak solution $\bu(\bx,\,t)$ of the three-dimensional Navier-Stokes equations 
\bel{ns1}
(\partial_{t} + \bu\cdot\nabla)\bu = \nu \Delta\bu - \nabla p + \bdf(\bx)\,,
\ee
with $\mbox{div}\,\bu = 0$, is called regular if the $H_{1}$-norm is continuous \cite{Leray34}. 
What is commonly referred to as `conditional regularity' can be achieved if it is found necessary 
to impose assumptions on certain system variables such as the velocity field. The early work of 
Prodi \cite{Prodi59}, Serrin \cite{Serrin63} and 
Ladyzhenskaya \cite{Lady69} can be summarized thus \cite{CF88,FMRT01}\,: every Leray-Hopf 
solution of the incompressible three-dimensional Navier-Stokes equations with $\bu \in 
L^{r}\left( (0,\, T)\,; L^{s}\right)$ is regular on $(0,\, T ]$ provided $2/r + 3/s = 1$, 
with $s \in (3,\,\infty]$, or if $\bu \in L^{\infty}\left((0,\, T)\,; L^{p}\right)$ with 
$p > 3$. The long-standing case $s = 3$ was finally settled by von Wahl \cite{vonWahl83} 
and Giga \cite{Giga86} who proved regularity in the space $C\!\left((0,\, T]\,; L^{3}\right)$\,: 
see also Kozono and Sohr \cite{KS97} and Escauriaza, Seregin and Sver\'ak \cite{ESS03}. 
In summary, the $s=3$ case seems tantalizingly close to the bounded case $s=2$, but not 
quite close enough.  More recent results exist where conditions are imposed on either the 
pressure or on one derivative of the velocity field\,: see the references in Kukavica and Ziane 
\cite{KZ06,KZ07}, Zhou \cite{YZ02}, Cao \& Titi \cite{CT08,CT10}, Cao \cite{Cao10}, Cao, 
Qin and Titi (for channel flows) \cite{CQT08}, Chen and Gala \cite{CG11} and the review by 
Doering \cite{CRDrev09}. Results on the direction of vorticity can be found in Constantin and 
Fefferman \cite{CF93} and Vasseur \cite{AVass08}, and those on the use of Besov spaces in 
Cheskidov and Shvydkoy \cite{CS11}). 

Finally, in recent work, Biswas and Foias \cite{BF12} have considered analyticity properties of 
Navier-Stokes solutions in which they have studied the maximal space analyticity radius associated 
with a regular solution involving Gevrey-class norms. Intermittency properties have also been 
studied by Grujic \cite{Grujic12} and Dascaliuc and Grujic \cite{DasGrujic12A,DasGrujic12B} 
using methods very different from those employed in this paper. 

\subsection{Motivation and notation}\label{mot}

In an entirely different thread of intellectual endeavour, the seminal experimental paper of 
Batchelor and Townsend  \cite{BT49} investigated the phenomenon of intermittency in wind 
tunnel turbulence by observing that the flatness of their signals (the ratio of the fourth order 
moment to the square of the second order moment) took much higher values than those 
expected for random Gaussian behaviour. They concluded that the vorticity is neither spatially 
nor temporally distributed in an even fashion but undergoes local clustering or spottiness, 
which is consistent with the appearance of spikes in the signals interspersed by longer quiescent 
periods.  This is now considered to be a classic characteristic of intermittency. These ideas have 
been developed and extended in many subsequent experiments and computations\,: 
see the papers by Kuo and Corrsin \cite{KC71}, Sreenivasan \cite{Sreenivasan85}, Meneveau 
and Sreenivasan \cite{MS91} and the books by Frisch \cite{Frisch95} and Davidson 
\cite{Davidson04} for further references. 

Most of these discussions have been based around Kolmogorov's statistical theory with the 
widespread use of velocity structure functions to study intermittent behaviour. However, structure 
functions are not easily translatable into results in Navier-Stokes analysis (see some of the arguments 
in Kuksin \cite{SK99} and Dascaliuc and Grujic \cite{DasGrujic12B}). The main difficulty lies in 
translating the special conditions needed to prove regularity listed in \S\ref{hist} into sensible physics 
while conversely making sense of the experimental observations in terms of Navier-Stokes variables. 
The two threads can be merged if the spiky nature of the vorticity field is considered in the context 
of $L^p$-norms of the vorticity $\bom = \mbox{curl}\,\bu$ with $p=2m$ 
\bel{Omdef}
\Omega_{m}(t) = \left(L^{-3}\I |\omega|^{2m}\,dV\right)^{1/2m} + \varpi_{0}\,,
\qquad\qquad m \geq 1\,,
\ee
on a periodic box $[0,\,L]^3$. The additive frequency $\varpi_{0} = \nu L^{-2}$ is present for 
technical reasons.  Clearly the $\Omega_{m}(t)$ are ordered for all $t$ such that
\bel{Omorder}
\varpi_{0} \leq \Omega_{1}(t) \leq \Omega_{2}(t) \leq \ldots \leq \Omega_{m}(t) 
\leq \Omega_{m+1}(t)\ldots \,,
\ee
where $L^{3/2}\Omega_{1}(t)$ is the $H_{1}$-norm. Control from above 
on any one of the $\Omega_{m}$ will also control the $H_{1}$-norm from above which is the 
ultimate key to regularity.  
\par\medskip
The Navier-Stokes equations have a well-known invariance property under the transformations 
$x' = \epsilon x$, $t' = \epsilon^{2}t$,  $u = \epsilon u'$ and $p = \epsilon^{2} p'$. Under these 
transformations $\Omega_{m}$ scales as
\beq{resc1}
\Omega_{m}^{\alpha_{m}}  = \epsilon\Omega_{m}^{'\alpha_{m}}\,,\qquad\qquad
\alpha_{m} = \frac{2m}{4m-3}\,.
\eeq
Thus it is natural to define 
\bel{Dmdef}
D_{m}(t) = \left[\varpi_{0}^{-1}\Omega_{m}(t)\right]^{\alpha_{m}}\,.
\ee
While the norms $\Omega_{m}$ are ordered with 
increasing $m$ as in (\ref{Omorder}), the $\alpha_{m}$ \textit{decrease} with $m$. Thus there is 
no natural ordering among the $D_{m}$.  Note also that $D_{1}$ is the square of the $H_{1}$-norm.
\par\medskip
For the forced case, the dimensionless Grashof number $Gr$ is based on the bounded-ness of the 
root-mean-square $f_{rms}^{2} = L^{-3}\|\bdf\|_{2}^{2}$ of the divergence-free 
forcing $\bdf(\bx)$ and is defined as 
\bel{Grdef}
Gr = \frac{L^{3}f_{rms}}{\nu^{2}}\,.
\ee

\subsection{Two fundamental results}\label{funda}

There are two results that form the basis of those given in later sections. The first is a theorem on time 
integrals or averages of $D_{m}$ \cite{JDGCMS11} which uses a result of Foias, Guillop\'e and Temam 
\cite{FGT}.  This proof will not be repeated\,:
\begin{theorem}\label{Dmav}
For $1 \leq m \leq \infty$, and $\alpha_{m}$ defined as $\alpha_{m} = \frac{2m}{4m-3}$, weak solutions 
obey
\bel{Ombd1}
\int_{0}^{t}D_{m}\,d\tau \leq c\left(t\,Gr^{2} + \eta_{1}\right)\,,
\ee
where $\eta_{1} = L\nu^{-3}E_{0}$ and $E_{0}$ is the initial energy. In the unforced case 
the right hand side is just $c\,\eta_{1}$. 
\end{theorem}
In \cite{JDGCMS11} this result was converted into a set of length scales. Let the time average 
up to time $T$ be defined by 
\bel{timeavdef}
\left< D_{m}\right>_{T} = \limsup_{D(0)}\frac{1}{T}\int_{0}^{T}D_{m}(\tau)\,d\tau
\ee
in which case (\ref{Ombd1}) can be re-expressed as
\bel{Dmbd}
\left< D_{m}\right>_{T} \leq c\,Gr^{2} + O\left(T^{-1}\right)\,.
\ee
Then, motivated by the definition of the Kolmogorov length for $m=1$, a set of length scales can 
be defined thus\,:
\bel{lamdef1}
\left(L\lambda_{m}^{-1}\right)^{2\alpha_{m}} := \left< D_{m}\right>_{T}\,.
\ee
In \cite{JDGCMS11} the bounds in (\ref{Dmbd}) were expressed as $Re^{3}$ instead of $Gr^{2}$ 
based on a device of Doering and Foias \cite{DF02} who used the square of the averaged velocity $U_{0}^{2} 
= L^{-3}\left<\|\bu\|_{2}^{2}\right>_{T}$ to define the Reynolds number $Re = U_{0}L\nu^{-1}$\,; 
for Navier-Stokes solutions this leads to the inequality $Gr \leq c\,Re^{2}$. However, the preference in this 
paper is to remain with the Grashof number $Gr$. In terms of $Re$ equation (\ref{lamdef1}) becomes 
\bel{lamdef2}
L\lambda_{m}^{-1} \leq c\,Re^{\frac{3}{2\alpha_{m}}}\,.
\ee
When $m=1$, $\alpha_{1}=2$, and thus $L\lambda_{1}^{-1} \leq c\,Re^{3/4}$, which is consistent 
with Kolmorgorov's statistical theory \cite{Frisch95}.
\par\medskip
The second result is a differential inequality for the $D_{m}$. Any attempt to time-differentiate the vorticity field 
creates problems because only weak solutions exist. Circumvention of this difficulty requires a contradiction 
strategy commonly used in geometric analysis\,: assume that there is a maximal interval of existence and 
uniqueness $[0,\,T^{*})$ which, for the three-dimensional Navier-Stokes equations, implies that $H_{1}(T^{*}) 
= \infty$. In any subsequent calculation, if the $H_{1}$-norm were to turn out bounded in the limit $t\to T^{*}$, 
then a contradiction would result and so the interval $[0,\,T^{*})$ could not be maximal.  Moreover, it cannot be 
zero, so $T^{*}$ would have to be infinite. 
\par\medskip
Define three frequencies 
\bel{varpidef}
\varpi_{1,m} = \varpi_{0}\alpha_{m}c_{1,m}^{-1}\qquad
\varpi_{2,m} = \varpi_{0}\alpha_{m}c_{2,m}\qquad
\varpi_{3,m} = \varpi_{0}\alpha_{m}c_{3,m}\,,
\ee
where the constants $c_{n,m}$ ($n = 1,2,3$) are algebraically increasing with $m$. 
The proof of the following theorem requires some variations on a previous result 
\cite{JDG10} and is relegated to Appendix \ref{appA}. The dot represents differentiation 
with respect to time\,:
\begin{theorem}\label{Dmthm}
For $1 \leq m < \infty$ on $[0,\,T^{*}]$ the $D_{m}(t)$ formally satisfy the set of inequalities 
\bel{Dminequal}
\dot{D}_{m} \leq D_{m}^{3}
\left\{-  \varpi_{1,m}\left(\frac{D_{m+1}}{D_{m}}\right)^{\rho_{m}} + \varpi_{2,m}\right\}\,,
\ee
where $\rho_{m} = \frac{2}{3} m(4m+1)$.  In the forced case there is an additive term 
$\varpi_{3,m}\Gr D_{m}$.
\end{theorem}

\rem{
The first integration in \S\ref{1stcond}, more appropriate for the unforced case, produces a 
conditional regularity result based on the relative sizes of $\int_{0}^{t}D_{m}d\tau$ and 
$\int_{0}^{t}D_{m+1}d\tau$ and . The result in \S\ref{2ndcond}, more appropriate for the 
forced case, is based on an assumed \textit{lower} bound on $\int_{0}^{t}D_{m+1}d\tau$ 
in terms of the Grashof number $Gr$. Both cases take advantage of the bounded time 
integrals of $D_{m}$ from Theorem \ref{Dmav}.}

\section{A conditional regularity result for unforced Navier-Stokes}\label{1stcond}

\subsection{Integration of the $D_{m}$ inequality}\label{1stinteg}

Theorem \ref{Dmthm} leads to the conclusion that solutions come under control 
pointwise in $t$ provided $D_{m+1}(t) \geq c_{\rho_{m}} D_{m}(t)$, where $c_{\rho_{m}} =  
\left[c_{1,m}c_{2,m}\right]^{1/\rho_{m}}$.  The following lemma shows that a time integral 
version of this controls solutions\,:
\begin{lemma}\label{3rdthm}
For any value of $1\leq m < \infty$ and $\varepsilon$ uniform in the range $0 < \varepsilon < 2$, if 
the integral condition is satisfied
\bel{3rdthmA}
\int_{0}^{t}D_{m+1}^{\varepsilon}d\tau \geq  c_{\varepsilon,\rho_{m}} \int_{0}^{t}D_{m}^{\varepsilon}d\tau
\qquad\qquad c_{\varepsilon,\rho_{m}} =  
\left[c_{1,m}c_{2,m}\right]^{\varepsilon/\rho_{m}}
\ee
then $D_{m}(t)$ obeys $D_{m}(t) \leq D_{m}(0)$ on the interval $[0,\,t]$. 
\end{lemma}
\textbf{Remark\,:} The case $\varepsilon = 1$ turns (\ref{3rdthmA}) into $\int_{0}^{t}D_{m+1}\,d\tau \geq  
c_{1,\rho_{m}} \int_{0}^{t}D_{m}\,d\tau$ both sides of which are bounded above. However, the fact that 
$\varepsilon$ can take small values suggests a logarithmic result which appears in the following theorem. The 
proof of lemma \ref{3rdthm} is included within its proof. 
\begin{theorem}
For any value of $1\leq m < \infty$, if the integral condition is satisfied
\bel{4ththmA}
\int_{0}^{t}\ln \left(\frac{1 + Z_{m}}{c_{4,m}}\right)\,d\tau \geq  0\,,\qquad\qquad Z_{m} = D_{m+1}/D_{m}
\ee
with $c_{4,m} = \left[ 2^{\rho_{m}-1}\left(1 + c_{1,m}c_{2,m}\right)\right]^{ \rho_{m}^{-1}}$, 
then $D_{m}(t) \leq D_{m}(0)$ on the interval $[0,\,t]$. 
\end{theorem}
\textbf{Remark 1\,:} This result may serve as an alternative to the Beale-Kato-Majda theorem \cite{BKM}.
\par\vspace{2mm}\noindent
\textbf{Remark 2\,:} The exponent $\rho_{m}^{-1}$ pulls $c_{4,m}$ down close to 2 for large $m$ which 
indicates that there needs to be enough intervals of time on which $Z_{m} > 1$ for (\ref{4ththmA}) to hold.
\par\vspace{2mm}\noindent
\textbf{Proof\,:} The proof of Lemma \ref{3rdthm} is addressed first. Divide (\ref{Dminequal}) by 
$D_{m}^{3-\varepsilon}$ and integrate to obtain 
\bel{H1}
[D_{m}(t)]^{\varepsilon-2} -  [D_{m}(0)]^{\varepsilon-2} \geq 
\varpi_{1,m}^{(\varepsilon)}\int_{0}^{t}\left(\frac{D_{m+1}}{D_{m}}\right)^{\rho_{m}}D_{m}^{\varepsilon}d\tau - 
\varpi_{2,m}^{(\varepsilon)}\int_{0}^{t}D_{m}^{\varepsilon}d\tau\,,
\ee
where $\varpi_{n,m}^{(\varepsilon)} = (2-\varepsilon) \varpi_{n,m}$.  Noting that $\rho_{m} \geq 10/3$, a H\"older 
inequality then easily shows that 
\beq{H2}
\int_{0}^{t} D_{m+1}^{\varepsilon}\,d\tau 
&=& 
\int_{0}^{t}\left[
\left(\frac{D_{m+1}}{D_{m}}\right)^{\rho_{m}}D_{m}^{\varepsilon}\right]^{\frac{\varepsilon}{\rho_{m}}}
\left[D_{m}^{\varepsilon}\right]^{\frac{\rho_{m}-\varepsilon}{\rho_{m}}}\,d\tau\non\\ 
&\leq &
\left(\int_{0}^{t}\left(\frac{D_{m+1}}{D_{m}}\right)^{\rho_{m}}D_{m}^{\varepsilon}d\tau\right)^{\frac{\varepsilon}{\rho_{m}}}
\left(\int_{0}^{t}D_{m}^{\varepsilon}d\tau\right)^{\frac{\rho_{m}-\varepsilon}{\rho_{m}}}\,.
\eeq
(\ref{H1}) can then be re-written as
\bel{H3}
D_{m}(t) \leq \left\{
[D_{m}(0)]^{\varepsilon-2} + 
\varpi_{1,m}^{(\varepsilon)}\frac{\left(\int_{0}^{t}D_{m+1}^{\varepsilon}d\tau\right)^{\rho_{m}/\varepsilon}}
{\left(\int_{0}^{t}D_{m}^{\varepsilon}d\tau\right)^{(\rho_{m}-\varepsilon)/\varepsilon}} - \varpi_{2,m}^{(\varepsilon)}\int_{0}^{t}D_{m}^{\varepsilon}\,d\tau
\right\}^{-\frac{1}{2-\varepsilon}}\,.
\ee
It is clear that no sign change can occur in the denominator of (\ref{H3}) if (\ref{3rdthmA}) holds.
\par\medskip
The proof of Theorem \ref{4ththmA} is now addressed. Divide (\ref{Dminequal}) by $D_{m}^{3}$ and 
integrate to obtain 
\beq{form2}
\frac{1}{2}\left([D_{m}(t)]^{-2} -  [D_{m}(0)]^{-2}\right) & \geq &
\varpi_{1,m}\int_{0}^{t}\left\{\left[1+ Z_{m}^{\rho_{m}}\right] - 
\left(1+\frac{\varpi_{2,m}}{\varpi_{1,m}}\right)\right\}d\tau \non\\
& \geq &
\frac{\varpi_{1,m}}{ 2^{\rho_{m}-1}}\int_{0}^{t}\left\{\left[1 + Z_{m}\right]^{\rho_{m}} - 
 2^{\rho_{m}-1}\left(1 + c_{1,m}c_{2,m}\right)\right\}d\tau\,,
\eeq
where we have used $\left(1 + Z_{m}\right)^{\rho_{m}} \leq 2^{\rho_{m}-1}\left(1 + Z_{m}^{\rho_{m}}\right)$. 
Re-arranging \& using Jensen's inequality
\bel{Jenson}
\frac{1}{t}\int_{0}^{t}\exp F(\tau)\,d\tau \geq \exp\left(\frac{1}{t}\int_{0}^{t}F(\tau)\,d\tau\right)\,,
\ee
 with $F = \rho_{m}\ln (1 + Z_{m})$, the RHS of (\ref{form2}) can be written as
\beq{form3}
\frac{1}{t}
\int_{0}^{t}\left\{\left[1 + Z_{m}\right]^{\rho_{m}} -  2^{\rho_{m}-1}\left(1 + c_{1,m}c_{2,m}\right)\right\}d\tau
& = & \frac{1}{t}\int_{0}^{t}\left\{\exp\left[\rho_{m}\ln (1 + Z_{m})\right] -  
2^{\rho_{m}-1}\left(1 + c_{1,m}c_{2,m}\right)\right\}d\tau\non\\
&\geq &
\exp\left[\frac{\rho_{m}}{t}\int_{0}^{t}\ln (1 + Z_{m})\,d\tau\right] - 
\exp\left[\rho_{m}\ln c_{4,m}\right]
\eeq
and thus no zero can develop if (\ref{4ththmA}) holds, as advertised.
\par\smallskip
In both cases if a zero cannot appear in the respective denominators then $D_{m}(t) \leq D_{m}(0)$. It follows that 
$\Omega_{m}(t)$ is bounded above and thus so is the $H_{1}$-norm ($\Omega_{1}$). \hfill $\blacksquare$

\section{Second integration of the $D_{m}$ inequality}\label{2ndcond}

\subsection{A lower bound on $\int_{0}^{t}D_{m}d\tau$}\label{lb}

Inclusion of the forcing in (\ref{Dminequal}) modifies it to 
\bel{DminequalG}
\dot{D}_{m} \leq D_{m}^{3}\left\{- \frac{1}{\varpi_{1,m}}\left(\frac{D_{m+1}}{D_{m}}\right)^{\rho_{m}} 
+ \varpi_{2,m}\right\} + \varpi_{3,m}\Gr D_{m}\,,
\ee
where $\rho_{m} = \frac{2}{3} m(4m+1)$.  To proceed, divide by $D_{m}^{3}$ (the case $\varepsilon =0$ of Theorem 
\ref{3rdthm}) to write (\ref{DminequalG}) as
\bel{2ndA}
\frac{1}{2}\frac{d~}{dt}\left(D_{m}^{-2}\right) \geq X_{m}\left(D_{m}^{-2}\right) - \varpi_{2,m}
\ee
where
\bel{2ndB}
X_{m} = \varpi_{1,m}\left(\frac{D_{m+1}}{D_{m}}\right)^{\rho_{m}}D_{m}^{2} - \varpi_{3,m}Gr\,.
\ee
A lower bound for $\int_{0}^{t}X_{m}d\tau$ can be estimated thus\,:
\beq{2ndE}
\int_{0}^{t} D_{m+1}\,d\tau 
&=& 
\int_{0}^{t}\left[
\left(\frac{D_{m+1}}{D_{m}}\right)^{\rho_{m}}D_{m}^{2}\right]^{\frac{1}{\rho_{m}}}D_{m}^{\frac{\rho_{m} - 2}{\rho_{m}}}
\,d\tau\non\\ 
&\leq &
\left(\int_{0}^{t}\left(\frac{D_{m+1}}{D_{m}}\right)^{\rho_{m}}D_{m}^{2}\,d\tau\right)^{\frac{1}{\rho_{m}}}
\left(\int_{0}^{t}D_{m}\,d\tau\right)^{\frac{\rho_{m}-2}{\rho_{m}}}t^{1/\rho_{m}}
\eeq
and so
\bel{2ndF}
\int_{0}^{t}X_{m}d\tau \geq \varpi_{1,m}
t^{-1}\frac{\left(\int_{0}^{t}D_{m+1}\,d\tau\right)^{\rho_{m}}}
{\left(\int_{0}^{t}D_{m}\,d\tau\right)^{\rho_{m}-2}}  - \varpi_{3,m}t\,Gr\,.
\ee
(\ref{2ndA}) integrates to
\bel{2ndD}
 \left[D_{m}(t)\right]^{2} \leq \frac{\exp\left\{-2\int_{0}^{t}X_{m}\,d\tau\right\}}
{\left[D_{m}(0)\right]^{-2} 
- 2\varpi_{2,m}\int_{0}^{t}\exp\left\{-2\int_{0}^{\tau}X_{m}\,d\tau'\right\}d\tau}\,.
\ee
Let us recall that $\rho_{m} = \frac{2}{3} m(4m+1)$ and let us also define
\bel{gamdef}
\gamma_{m} = \frac{\alpha_{m+1}}{2\left(m^{2}-1\right)}\,,
\ee
then
\par\medskip\noindent
\begin{theorem}\label{5ththm}
On  the interval $[0,\,t]$ if there exists a value of $m$ lying in the range $1 < m < \infty$, with 
initial data $\left[D_{m}(0)\right]^{2} < C_{m}Gr^{\Delta_{m}}$, for which the integral lies 
on or above the critical value
\bel{5thm1}
c_{m}\left(t\,Gr^{2\delta_{m}} + \eta_{2}\right) \leq \int_{0}^{t} D_{m}\,d\tau
\ee
where $\eta_{2} \geq \eta_{1}Gr^{2(\delta_{m}-1)}$ and where $(1 \leq \Delta_{m} \leq 4)$
\bel{5thm2}
\Delta_{m} =  4\left\{\delta_{m}(1+\rho_{m}\gamma_{m}) - \rho_{m}\gamma_{m}\right\}\qquad\mbox{with}\qquad
\frac{1 + 4\rho_{m}\gamma_{m}}{4(1+\rho_{m}\gamma_{m})} < \delta_{m} < 1\,,
\ee
then $D_{m}(t)$ decays exponentially on $[0,\,t]$.
\end{theorem}
\par\medskip\noindent
\textbf{Remark\,:} $\delta_{m} \searrow 11/20$ for large $m$ so enough slack lies between the upper and lower bounds 
on $\int_{0}^{t}D_{m}\,d\tau$.
\par\medskip\noindent
\textbf{Proof\,:} It is not difficult to prove that $\Omega_{m}^{m^{2}} \leq \Omega_{m+1}^{\,m^{2}-1}\Omega_{1}$
for $m > 1$, from which it is easily found that
\bel{newthm2}
\frac{\int_{0}^{t}D_{m+1}\,d\tau}{\int_{0}^{t}D_{m}\,d\tau} \geq 
\left(\frac{\int_{0}^{t}D_{m}\,d\tau}{\int_{0}^{t}D_{1}\,d\tau}\right)^{\gamma_{m}}\,.
\ee
Thus (\ref{2ndF}) can be re-written as
\beq{newthm3}
\int_{0}^{t}X_{m}d\tau 
&\geq& \varpi_{1,m}t^{-1} 
\frac{\left(\int_{0}^{t}D_{m}\,d\tau\right)^{\rho_{m}\gamma_{m}+2}}{\left(\int_{0}^{t}D_{1}\,d\tau\right)^{\rho_{m}\gamma_{m}}}
 - \varpi_{3,m}t\,Gr\non\\
&\geq& c_{m}t\left(\varpi_{1,m} Gr^{\Delta_{m}} - \varpi_{3,m}Gr\right)
\eeq
having used the assumed lower bound in the theorem and the upper bound of $\int_{0}^{t}D_{1}\,d\tau$.
Moreover, to have the dissipation greater than forcing requires $\Delta_{m} > 1$ so $\delta_{m}$ must lie 
in the range as in (\ref{5thm2}) because $1 < \Delta_{m} \leq 4$. 
For large $Gr$ the negative $Gr$-term in (\ref{2ndD}) is dropped so the integral in 
the denominator of (\ref{2ndD}) is estimated as
\beq{newthm5}
\int_{0}^{t}\exp\left(-2\int_{0}^{\tau}X_{m}d\tau'\right)\,d\tau 
\leq \left[2\tilde{c}_{m}\varpi_{1,m}\right]^{-1}Gr^{-\Delta_{m}}
\left(1 - \exp\left[- 2\varpi_{1,m}\tilde{c}_{m}t\, Gr^{\Delta_{m}}\right]\right)\,,
\eeq
and so the denominator of (\ref{2ndD}) satisfies
\bel{newthm6}
\mbox{Denominator} \geq \left[D_{m}(0)\right]^{-2} - c_{2,m}c_{1,m}\left(2\tilde{c}_{m}\right)^{-1}
Gr^{-\Delta_{m}}\left(1 - \exp\left[- 2\varpi_{1,m}\tilde{c}_{m}t\,Gr^{\Delta_{m}}\right]\right)\,.
\ee
This can never go negative if $ \left[D_{m}(0)\right]^{-2} >  
c_{1,m}c_{2,m}\left(2\tilde{c}_{m}\right)^{-1}Gr^{-\Delta_{m}}$, 
which means $D_{m}(0) < C_{m}Gr^{\shalf\Delta_{m}}$.  \hfill $\blacksquare$

\subsection{A mechanism for intermittency}\label{inter} 

A major feature of intermittent flows lies in the strong, spiky, excursions of the vorticity away from 
averages with periods of relative inactivity between the spikes. How do these aperiodic cycles appear 
in solutions and does the critical lower bound imposed as an assumption in Theorem \ref{5ththm} 
lead to this?  Using the average notation $\left<\cdot\right>_{t}$, equation (\ref{2ndD}) shows that 
if $\left<D_{m}\right>_{t}$ lies above critical then $D_{m}(t)$ collapses exponentially. Experimentally, 
signals go through cycles of growth and collapse so it is not realistic to expect this 
critical lower bound to hold for all time. 
.\par\vspace{-27mm}
\bc
\begin{minipage}[htb]{9cm}
\setlength{\unitlength}{.9cm}
\begin{picture}(11,11)(0,0)
\put(0,0){\vector(0,1){8}}
\multiput(0,0)(0.1,0){100}{.}
\put(-0.75,7.5){\makebox(0,0)[b]{$D_{m}(t)$}}
\put(-0.5,3){\makebox(0,0)[b]{$Gr^{2}$}}
\put(2.5,3.1){\makebox(0,0)[b]{\tiny~~~~~~~~~~~~~~~~~~~~~~upper~bound~of~$\left<D_{m}\right>_{t}$}}
\put(3,1.6){\makebox(0,0)[b]{\tiny~~~~~~~~~~~~~~critical~lower-bound~of~$\left<D_{m}\right>_{t}$}}
\put(-.6,1.5){\makebox(0,0)[b]{$Gr^{2\delta_{m}}$}}
\put(10.3,-.1){\makebox(0,0)[b]{$t$}}

\multiput(6,0)(0,.1){50}{.}
\multiput(8,0)(0,.1){60}{.}
%
\multiput(0,3)(.1,0){92}{.}
\multiput(0,1.5)(.1,0){92}{.}
\put(0,-.5){\makebox(0,0)[b]{\small $t_{0}$}}
\put(6.1,-.5){\makebox(0,0)[b]{\small $t_{1}$}}
\put(8.1,-.5){\makebox(0,0)[b]{\small $t_{2}$}}
\put(7.1,1){\makebox(0,0)[b]{\tiny potential singularities}}
\put(7.1,.7){\makebox(0,0)[b]{$\uparrow$}}
\put(6.7,.3){\makebox(0,0)[b]{$\uparrow$}}
\qbezier[500](0.01,8)(-.1,-0.25)(6,0.05)
\qbezier[500](6,0.05)(8,0)(8,6)
\qbezier[1000](8.05,6)(8,0)(10,0)
\end{picture}
\end{minipage}
\ec
\par\vspace{3mm}
\bc
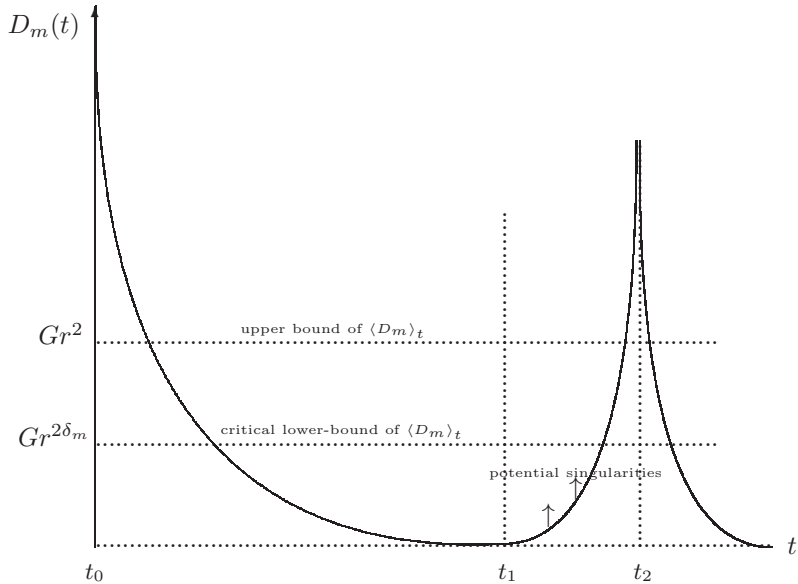
{\textbf{Figure 1\,:} \small A cartoon of $D_{m}(t)$ versus $t$ illustrating 
the four phases of intermittency.\\ The vertical arrows depict the region where there is 
the potential for needle-like singular behaviour.}
\ec

\par\vspace{2mm}\noindent
To understand intermittency we turn to Figure 1 and draw the horizontal line at $Gr^{2\delta_{m}}$ 
as the critical lower bound on $\left<D_{m}\right>_{t}$. Within this allowed range, $D_{m}(t)$ will 
decay exponentially fast. Because integrals must take account of history, there will be a delay 
before $\left<D_{m}\right>_{t}$ decreases below the value above which a zero in the denominator 
of (\ref{2ndD}) can be prevented (at $t_{1}$)\,: at this point all constraints are removed and the 
pointwise solution $D_{m}(t)$ is free to grow rapidly again in the interval $t_{1} \leq t \leq t_{2}$. 
If the value of this integral drops below critical then it is in this interval that the occurrence of 
singular events (depicted by vertical arrows)) must still formally be considered\,: see the discussion 
in \S\ref{concl-sect}. Provided a solution still exists at this point, growth in $D_{m}$ will be such that, 
after another delay, it will force $\left<D_{m}\right>_{t}$ above critical and the system, with a re-set 
of initial conditions at $t_{2}$, is free to go through another cycle akin to a relaxation oscillator. 

\section{Conclusion}\label{concl-sect}

The results of Theorems \ref{4ththmA} and \ref{5ththm} are new. More appropriate for the unforced 
case, Theorem \ref{4ththmA} can be summarized thus\,: if there sufficient regions for which $Z_{m} > 1$ 
on the $t$-axis, such that 
\bel{ccl0}
\int_{0}^{t} \ln\left(\frac{1+ Z_{m}}{c_{4,m}}\right)\,d\tau \geq 0\qquad\qquad Z_{m} = D_{m+1}/D_{m}
\ee
then $D_{m}(t) \leq D_{m}(0)$. In fact both of the regimes  $Z_{m} \lessgtr 1$ are physically realistic but 
numerical experiments may suggest which of these two regimes are the most manifest and whether a  
cross-over from one to the other occurs.
\par\medskip
For the forced case, Theorem \ref{5ththm} can be summarized thus\,: if
\bel{ccl1}
\int_{0}^{t}D_{m}\,d\tau \geq c_{m}\left(t\,Gr^{2\delta_{m}} + \eta_{2}\right)
\ee
then $D_{m}(t)$ collapses exponentially. The main question is the meaning of the regime below this critical value
\bel{ccl2}
\int_{0}^{t}D_{m}\,d\tau < c_{m}\left(t\,Gr^{2\delta_{m}} + \eta_{2}\right)\,.
\ee
It might be natural to suppose, on intuitive grounds, that singular behaviour would be less likely to occur when 
the upper bound is smaller, yet this behaviour cannot be wholly ruled out. What can be ruled out are potentially 
singular spikes that substantially contribute to the time integral of $D_{m}(t)$ because they would push it 
over its critical value, thereby forcing exponential collapse in $D_{m}(t)$. However, there still remains the 
possibility of needle-like singular spikes (depicted by arrows in Figure 1) that contribute little or nothing 
to the time integral of $D_{m}(t)$ in (\ref{ccl2}). It has been shown that the time-axis can potentially
be divided into `good' and `bad' intervals, the name of this second set implying that no control over
solutions has yet been found \cite{JDG10}. To summarize the argument in reference \cite{JDG10}, it is very easy 
to show that for an arbitrary set of parameters $0 < \mu_{m} < 1$, 
\bel{bad1}
\int_{0}^{t}\left(
\left[\frac{D_{m+1}}{D_{m}}\right]^{\frac{1-\mu_{m}}{\mu_{m}}} - 
\left[c_{m}^{-1}Gr^{-2}D_{m+1}^{\mu_{m}}\right]^{\frac{1-\mu_{m}}{\mu_{m}}}
\right)\,d\tau\geq 0\,.
\ee
Thus there are potentially `bad' intervals of the $t$-axis on which 
\bel{bad2}
\frac{D_{m+1}}{D_{m}} \geq c_{m}^{-1}Gr^{-2}D_{m+1}^{\mu_{m}}
\ee
but on which no upper bounds have been found. 
Using the fact that $\Omega_{m+1} \geq \Omega_{m}$, it follows that on these intervals 
\bel{bad3}
D_{m+1} \geq c_{m} Gr^{\frac{2}{1  + \mu_{m} - \alpha_{m+1}/\alpha_{m}}}\,.
\ee 
For large $m$ reduces to $D_{m+1} \geq c_{m} Gr^{2/\mu_{m}}$. Given that $\mu_{m}$ could be chosen
very small these lower bounds could be very large indeed, incidentally too large for the Navier-Stokes
equations to remain valid. It is possible that these are the root cause of the potential singularities 
discussed above and labelled by vertical arrows in Figure 1.

It is also possible to interpret this behaviour informally using the so-called $\beta$-model of Frisch, Sulem and 
Nelkin \cite{FSN78} who modelled a Richardson cascade by taking a vortex of scale $\ell_{0} \equiv L$ which
cascades into daughter vortices, each of scale $\ell_{n}$. The vortex domain halves at each step\,: thus $\ell_{0}/
\ell_{n} = 2^{n}$. The self-similarity dimension $d$ is then introduced by considering the number of offspring 
at each step as $2^{d}$, where $d$ is formally allowed to take non-integer values. In $d$ dimensions the  
Kolmogorov scaling calculations for velocity, turn-over time and other variables have multiplicative factors 
proportional to $\left(\ell_{0}/\ell_{n}\right)^{(3-d)/3}$\,: see \cite{Frisch95,FSN78}.  Equating the 
turn-over and viscous times in the standard manner one arrives at ($\ell_{d}$ is their viscous dissipation 
length)
\bel{Kold1}
L\lambda_{m}^{-1} \equiv \ell_{0}/\ell_{d} \sim Re^{\frac{3}{d+1}}\,.
\ee
This gives the Kolmogorov inverse scale of $Re^{3/4}$ in a three-dimensional domain\footnote{In 
\cite{JDGCMS11} this idea was used to compare (\ref{Kold1}) to the upper bound in (\ref{lamdef2}) to get 
$d_{m}+1  = 2\alpha_{m}$ so that $d_{m} = \frac{3}{4m-3}$, where an $m$-label was appended to $d$. 
This assigned a corresponding self-similarity dimension $d_{m}$ to lower-dimensional vortical structures that 
require values of $m > 1$ for their resolution.}. To interpret the meaning of (\ref{ccl2}) in the light of 
(\ref{lamdef1}) requires the conversion of (\ref{ccl3}) into Reynolds number notation ($Gr \to Re^{2}$) 
as in \cite{DF02}
\bel{ccl3}
\int_{0}^{t}D_{m}\,d\tau < c_{m}\left(t\,Re^{3\delta_{m}} + \eta_{2}\right)\,.
\ee
with corresponding length scales
\bel{ccl4}
L\lambda_{m}^{-1} \lesssim Re^{\frac{3\delta_{m}}{2\alpha_{m}}}\,.
\ee
This makes $d_{m} = 2\alpha_{m}\delta_{m}^{-1} - 1$ where the range of $\delta_{m}$ is given in (\ref{5thm2}).
In the large $m$ limit the range of $\delta_{m}$ widens to $11/20 < \delta_{m} < 1$ implying that $d_{m}$ lies in 
the range $0 < d_{m} < 9/11$. However, it is conceivable that the sharp result for the lower bound on $\delta_{m}$ 
is $1/2$ which alters the range to $0 < d_{m} < 1$. The upper bound of, or near, unity is consistent with the result 
of Caffarelli, Kohn and Nirenberg \cite{CKN} who showed that the singular set of the three-dimensional Navier-Stokes 
equations in four-dimensional space-time has zero 1-dimensional Hausdorff measure. Thus, it is possible that singularities 
that make no contribute to the integral in (\ref{ccl3}) may conceivably be related to the CKN singular set. Whether or 
not these are physically realisable is open to question. 

\begin{acknowledgements} 
Thanks are due to  Robert Kerr (Warwick) and Darryl Holm (Imperial) for discussions on this 
problem and to Mihaela Ifrim (UC Davis) for a critical reading of the manuscript.
\end{acknowledgements}

\appendix

\section{Proof of Theorem \ref{Dmthm}}\label{appA}

The following proof has some important variations on that given in \cite{JDG10}. Consider 
$J_{m} = \I |\bom|^{2m}dV$ such that
\bel{Hm1}
\frac{1}{2m}\dot{J}_{m} = \I |\bom|^{2(m-1)}\bom\cdot
\left\{\nu\Delta\bom + \bom\cdot\nabla\bu + \hbox{curl}\bdf\right\}\,dV\,.
\ee 
Bounds on the three constituent parts of (\ref{Hm1}) are dealt with in turn, 
culminating in a differential inequality for $J_{m}$. 
\par\medskip\noindent
\textit{a)~The Laplacian term\,:} Let $\phi = \omega^{2} = \bom\cdot\bom$. Then
\beq{s1a}
\I |\bom|^{2(m-1)}\bom\cdot\Delta\bom\,dV 
&=&\I \phi^{m-1}\left\{\Delta\left(\frac{1}{2}\phi\right) - |\nabla\bom|^{2}\right\}\,dV\nonumber\\
&\leq& \I \phi^{m-1}\Delta\left(\frac{1}{2}\phi\right)\,dV\,.
\eeq
Using the fact that $\Delta(\phi^{m}) = m\left\{(m-1)\phi^{m-2}|\nabla\phi|^{2} + 
\phi^{m-1}\Delta\phi\right\}$ we obtain
\beq{s1ex1}
\I |\bom|^{2(m-1)}\bom\cdot\Delta\bom\,dV 
&\leq& -\frac{1}{2}(m-1)\I\phi^{m-2}|\nabla\phi|^{2}\,dV + 
\frac{1}{2m}\I\Delta(\phi^{m})\,dV\nonumber\\
&=& - \frac{2(m-1)}{m^2}\I|\nabla(\omega^{m})|^{2}\,dV\,,
\eeq
having used the Divergence Theorem. Thus we have 
\beq{s1ex2}
\I |\bom|^{2(m-1)}\bom\cdot\Delta\bom\,dV \leq \left\{
\begin{array}{cl}
-\I|\nabla\bom|^{2}]\,dV & m=1\,,\\
-\frac{2}{\tilde{c}_{1,m}}\I|\nabla A_{m}|^{2}\,dV & m\geq 2\,.
\end{array}
\right.
\eeq
where $A_{m}= \omega^{m}$ and $\tilde{c}_{1,m} = m^{2}/(m-1)$ with equality at $m=1$. 
The negativity of the right hand side of (\ref{s1ex2}) is important. Both 
$\|\nabla A_{m}\|_{2}$ and $\|A_{m}\|_{2}$ will appear later in the proof.  
\par\medskip\noindent
\textit{b)~The nonlinear term in (\ref{Hm1})\,:} After a H\"older inequality, the 
second term in (\ref{Hm1}) becomes
\beq{s2a}
\I |\bom|^{2m}|\nabla\bu|\,dV
&\leq& \left(\I |\nabla\bu|^{2(m+1)}dV\right)^{\frac{1}{2(m+1)}}
\left(\I |\bom|^{2(m+1)}dV\right)^{\frac{m}{2(m+1)}}
\left(\I |\bom|^{2m}dV\right)^{\frac{1}{2}}\non\\
&\leq& c_{m}\left(\I |\bom|^{2(m+1)}dV\right)^{\frac{1}{2}}
\left(\I |\bom|^{2m}dV\right)^{\frac{1}{2}}\non\\
&=& c_{m}J_{m+1}^{1/2}\,J_{m}^{1/2}\,,
\eeq
where the inequality $\|\nabla\bu\|_{p} \leq c_{p} \|\bom\|_{p}$ for $p \in (1,\,\infty)$ has been 
used, which is based on a Riesz transform\,: note the exclusion of the case $m=\infty$ where a logarithm
of norms of derivatives is necessary \cite{BKM} -- see \cite{Kukavica03} for remarks on 
$L^{\infty}$-estimates. Together with (\ref{s1a}) this makes (\ref{Hm1}) into 
\bel{Hm2}
\frac{1}{2m}\dot{J}_{m} \leq -\frac{\nu}{\tilde{c}_{1,m}}\I|\nabla(\omega^{m})|^{2}\,dV 
+ c_{m}J_{m+1}^{1/2}J_{m}^{1/2} + \I|\bom|^{2(m-1)}\bom\cdot\hbox{curl}\bdf\,dV\,.
\ee
\par\medskip\noindent
\textit{c)~The forcing term in (\ref{Hm1})\,:}  Now we use the smallest scale in the forcing $\ell$
with $\ell = L/2\pi$ to estimate the last term in (\ref{Hm2})
\beq{f2}
\I|\bom|^{2(m-1)}\bom\cdot\hbox{curl}\bdf\,dV &\leq& \|\bom\|_{2m}^{2m-1}\|\nabla\bdf\|_{2m}
\eeq
However, by going up to at least $N\geq 3$ derivatives in a Sobolev inequality and using our
restriction of single-scale forcing at $k\sim \ell^{-1}$ (with $\ell = L/2\pi$) it can easily be shown 
that $\|\nabla\bdf\|_{2m} \leq c\,\|\bdf\|_{2}\ell^{\frac{3-5m}{2m}}$ and so
\beq{f3}
\left|\I|\bom|^{2(m-1)}\bom\cdot\hbox{curl}\bdf\,dV\right|
&\leq& c\,\Omega_{m}^{2m-1}L^{3}\varpi_{0}^{2}\,Gr\,.
\eeq
\par\medskip\noindent
\textit{d)~A differential inequality for $J_{m}$\,:} Recalling that $A_{m} = \omega^{m}$ 
allows us to re-write $J_{m+1}$ as
\bel{s3a}
J_{m+1} = \|A_{m}\|_{2(m+1)/m}^{2(m+1)/m}\,.
\ee
A Gagliardo-Nirenberg inequality yields
\bel{s3b}
\|A_{m}\|_{2(m+1)/m} \leq c_{m}\,\|\nabla A_{m}\|_{2}^{3/2(m+1)}
\|A_{m}\|_{2}^{(2m-1)/2(m+1)}
\ee
which means that 
\bel{s3c}
J_{m+1} \leq c_{m}\left(\I|\nabla(\omega^{m})|^{2}\,dV\right)^{3/2m}J_{m}^{(2m-1)/2m}\,.
\ee
With the definition
\bel{betamdef}
\beta_{m} = \frac{4}{3}m(m+1) 
\ee
(the factor of $\frac{4}{3}$ is different from that in \cite{JDG10}), (\ref{s3c}) can be used to form $\Omega_{m+1}$ 
\beq{Jm1a}
\Omega_{m+1} = \left(L^{-3}J_{m+1} + \varpi_{0}^{2(m+1)}\right)^{1/2(m+1)} 
&\leq& c_{m}\left(L^{-1}\I|\nabla(\omega^{m})|^{2}\,dV + \varpi_{0}^{2m}\right)^{1/\beta_{m}}\non\\
&\times&\left[\left(L^{-3}J_{m}\right)^{1/2m} + \varpi_{0}\right]^{(2m-1)/2(m+1)}
\eeq
which converts to
\bel{Jm1b}
c_{m}\left(L^{-1}\I|\nabla(\omega^{m})|^{2}\,dV + \varpi_{0}^{2m}\right)
\geq \left(\frac{\Omega_{m+1}}{\Omega_{m}}\right)^{\beta_{m}}\Omega_{m}^{2m}\,.
\ee
This motivates us to re-write (\ref{Hm2}) as
\beq{Jm2}
\frac{1}{2m}\big(L^{-3}\dot{J}_{m}\big) \leq -\frac{\varpi_{0}}{\tilde{c}_{1,m}}
\left(L^{-1}\I|\nabla(\omega^{m})|^{2}\,dV\right) &+& \tilde{c}_{2,m}\left(L^{-3}J_{m+1}\right)^{1/2}
\Big(L^{-3}J_{m}\Big)^{1/2}\non\\
&+&  c_{3,m}\varpi_{0}^{2}\Omega_{m}^{2m-1}\Gr\,.
\eeq
Converting the $J_{m}$ into $\Omega_{m}$ and using the fact that $\Omega_{m} \geq \varpi_{0}$ 
\beq{Jm5}
\dot{\Omega}_{m} &\leq& \Omega_{m}\left\{-\frac{\varpi_{0}}{c_{4,m}}
\left(\frac{\Omega_{m+1}}{\Omega_{m}}\right)^{4m(m+1)/3} 
+ c_{5,m}\left(\frac{\Omega_{m+1}}{\Omega_{m}}\right)^{m+1}\Omega_{m} 
+ c_{6,m}\varpi_{0}\Gr\right\}
\eeq
Using a H\"older inequality on the central term on the right hand side, with the definition of $\beta_{m}$ given 
in (\ref{betamdef}), (\ref{Jm5}) finally becomes
\bel{new4}
\dot{\Omega}_{m} \leq \varpi_{0}\Omega_{m}\left\{-\frac{1}{c_{1,m}}
\left(\frac{\Omega_{m+1}}{\Omega_{m}}\right)^{\beta_{m}} 
+ c_{2,m}\left(\varpi_{0}^{-1}\Omega_{m}\right)^{2\alpha_{m}} + c_{3,m}\Gr\right\}\,.
\ee
Given the definition $D_{m} = \left(\varpi_{0}^{-1}\Omega_{m}\right)^{\alpha_{m}}$ 
it is found that 
\bel{new5}
\left(\frac{\Omega_{m+1}}{\Omega_{m}}\right)^{\beta_{m}} =
\left(\frac{D_{m+1}}{D_{m}}\right)^{\rho_{m}} D_{m}^{2}
\ee
having used the fact that 
\bel{new6}
\left(\frac{1}{\alpha_{m+1}} - \frac{1}{\alpha_{m}}\right)\beta_{m} = 2
\ee
and $\rho_{m} = \beta_{m}/\alpha_{m+1}$. This converts (\ref{new4}) to (\ref{Dminequal}) 
of the theorem with $m=\infty$ excluded. The constants $c_{n,m}$ for $n=1,2,3$ grow 
algebraically with $m$. \hfill $\blacksquare$


\end{document}